\documentclass[aip,jcp,preprint]{revtex4-1}
\usepackage{amsmath,amssymb,bm,graphicx,color,braket}

\usepackage{algpseudocode}

\newcommand{\alert}[1]{\textcolor{black}{#1}}

\newcommand{\al}{\alpha}
\newcommand{\be}{\beta}
\newcommand{\ga}{\gamma}
\newcommand{\de}{\delta}
\newcommand{\ze}{\zeta}
\newcommand{\G}{\Gamma}
\newcommand{\la}{\lambda}
\newcommand{\bk}{\mathbf{k}}
\newcommand{\br}{\mathbf{r}}
\newcommand{\bA}{\mathbf{A}}
\newcommand{\bB}{\mathbf{B}}
\newcommand{\bC}{\mathbf{C}}
\newcommand{\bD}{\mathbf{D}}
\newcommand{\bP}{\mathbf{P}}
\newcommand{\bQ}{\mathbf{Q}}
\newcommand{\bu}{\mathbf{u}}
\newcommand{\RD}{\mathbb{R}^D}
\newcommand{\SD}{\mathbb{S}^D}

\newcommand{\EHF}{E_{\rm HF}}
\newcommand{\Ecred}{\bar{E}_c}
\newcommand{\mEh}{m$E_{\rm h}$}
\newcommand{\nG}{N_{\rm G}}
\newcommand{\nY}{N_{\rm Y}}
\newcommand{\co}{\cos\omega}
\newcommand{\Geg}{C_n^\la(\co)}
\newcommand{\half}{\frac{1}{2}}

\newcommand{\Ga}{G_\al^\bA}
\newcommand{\Gb}{G_\be^\bB}
\newcommand{\Gc}{G_\ga^\bC}
\newcommand{\Gd}{G_\de^\bD}
\newcommand{\GGab}{\left( \Ga \vert \Gb \right)}
\newcommand{\GGcd}{\left( \Gc \vert \Gd \right)}
\newcommand{\GTG}{\left( \Ga \left| \hat{T} \right| \Gb \right)}
\newcommand{\GGGG}{\left( \Ga\Gb \vert \Gc\Gd \right)}
\newcommand{\ds}{\displaystyle}

\begin{document}

%==================================================
\title{Basis functions for electronic structure calculations on spheres}
%==================================================
\author{Peter M.W. Gill}
\email{peter.gill@anu.edu.au}
\author{Pierre-Fran\c{c}ois Loos}
\email{Corresponding author: pf.loos@anu.edu.au}
\author{Davids Agboola}
\affiliation{Research School of Chemistry, Australian National University, ACT 2601, Australia}

%==========
\begin{abstract}
We introduce a new basis function (the spherical gaussian) for electronic structure calculations on spheres of any dimension \alert{$D$}.  We find \alert{general} expressions for the one- and two-electron integrals and propose an efficient computational algorithm incorporating the Cauchy-Schwarz bound.  \alert{Using numerical calculations for the $D = 2$ case,} we show that spherical gaussians are more efficient than spherical harmonics when the electrons are strongly localized.
\end{abstract}
%==========

\maketitle

%===============
\section{Introduction}
%===============
Consider electronic structure calculations in which the electrons move in $D$-dimensional cartesian space $\RD$.  If the molecular orbitals (MOs) are delocalized throughout space, the plane waves\cite{Bloch28}
\begin{equation}
	q_\bk(\br) = \exp(i \bk \cdot \br),		\qquad		\br \in \RD
\end{equation}
form a good basis, because the product of two is a third
\begin{equation}
	q_{\bk_1}(\br) q_{\bk_2}(\br) = q_{\bk_1+\bk_2}(\br)
\end{equation}
If the MOs are localized, the cartesian gaussians\cite{Boys50}
\begin{equation} \label{eq:gal}
	g_\al^\bA(\br) = \exp(-\al |\br-\bA|^2),	\qquad		\br \in \RD
\end{equation}
are effective, again because the product of two is a third
\begin{gather}
	g_\al^\bA(\br) g_\be^\bB(\br) = K g_{\al+\be}^\bP(\br)	\\
	K = \exp(-\al\be |\bA-\bB|^2 / (\al+\be))					\\
	\bP = (\al \bA + \be \bB) / (\al+\be)
\end{gather}

Now consider calculations\cite{TEOAS09, QuasiExact09, ExSpherium10, Glomium11, UEGs12, QR12, Ringium13, gLDA14} in which the electrons move on the $D$-dimensional sphere $\SD$, \textit{i.e.} on the surface of a $(D+1)$-dimensional unit ball.  If the average interelectronic separation $r_s$ is small, the MOs are delocalized over the sphere and the (hyper)spherical harmonics\cite{AveryBook}
\begin{equation}
	Q_{k,K}(\br) = Y_{k,K}(\br),		\qquad		\br \in \SD
\end{equation}
(where $K$ is a composite index) provide a useful basis because it is single-valued and the product of two of these functions is a finite sum of several others
\begin{equation}
	Q_{k_1,K_1}(\br) Q_{k_2,K_2}(\br) = \sum_k \sum_K c_{k,K} Q_{k,K}(\br)
\end{equation}
where $c_{k,K}$ is a generalized Clebsch-Gordan coefficient.  However, if $r_s$ is large and the MOs are localized, what are good basis functions?

In this paper, we propose that spherical gaussian functions (SGFs) are a natural basis set for localized MOs on a sphere.  In Section \ref{sec:SGF}, we define SGFs and show that the product of two is a third.  In Section \ref{sec:ints}, we resolve the Coulomb operator on a sphere and use this to derive expressions for integrals over SGFs on the unit sphere.  Section \ref{sec:algo} discusses implementation details of our integral formulae and Section \ref{sec:results} presents some numerical results for Wigner molecules on a 2-sphere.  Atomic units are used throughout.

\begin{table*}	
	\caption{\label{tab:formulae}
		Overlap, kinetic and electron repulsion integrals\footnote
			{$i_n$ is a modified spherical Bessel function, $P_n$ is a Legendre polynomial, $T_n$ and
			$U_n$ are Chebyshev polynomials\cite{NISTbook} and $s_n = \sum_{p=1}^n (2p-1)^{-1}$ is a harmonic number.}
		over spherical gaussian functions (SGFs) on the unit $D$-sphere}
	\footnotesize
	\begin{ruledtabular}	
		\begin{tabular}{cccc}
			$D$&			$\ds \GGab$			&			$\ds \GTG / \GGab$		&		$\ds \GGGG / \GGab / \GGcd$			\\
			\hline
			1	&	$\ds \frac{I_0(\ze)}{\sqrt{I_0(2\al) I_0(2\be)}}$
				&	$\ds \frac{I_1(\ze)}{I_0(\ze)} \frac{\al\be\cos\theta}{2\ze} - \frac{I_2(\ze)}{I_0(\ze)}\frac{(\al\be\sin\theta)^2}{2\ze^2}$
				&	$\ds -\frac{4}{\pi} \sum_{n=1}^\infty \frac{I_n(\ze)}{I_0(\ze)} \frac{I_n(\eta)}{I_0(\eta)} s_n T_n(\cos\chi)$					\\
			2	&	$\ds \frac{i_0(\ze)}{\sqrt{i_0(2\al) i_0(2\be)}}$
				&	$\ds \frac{i_1(\ze)}{i_0(\ze)} \frac{\al\be\cos\theta}{\ze} - \frac{i_2(\ze)}{i_0(\ze)}\frac{(\al\be\sin\theta)^2}{2\ze^2}$
				&	$\ds \sum_{n=0}^\infty \frac{i_n(\ze)}{i_0(\ze)} \frac{i_n(\eta)}{i_0(\eta)} P_n(\cos\chi)$									\\
			3	&	$\ds \frac{I_1(\ze)/\ze}{\sqrt{I_1(2\al) I_1(2\be) / (4\al\be)}}$
				&	$\ds \frac{I_2(\ze)}{I_1(\ze)} \frac{3\al\be\cos\theta}{2\ze} - \frac{I_3(\ze)}{I_1(\ze)}\frac{(\al\be\sin\theta)^2}{2\ze^2}$
				&	$\ds \frac{2}{\pi} \sum_{n=1}^\infty \frac{I_n(\ze)}{I_1(\ze)} \frac{I_n(\eta)}{I_1(\eta)} \frac{nU_{n-1}(\cos\chi)}{n^2-1/4}$	\\
		\end{tabular}
	\end{ruledtabular}
\end{table*}

%=============================
\section{Spherical Gaussian Functions}
\label{sec:SGF}
\begin{figure}
	\includegraphics[width=0.44\textwidth]{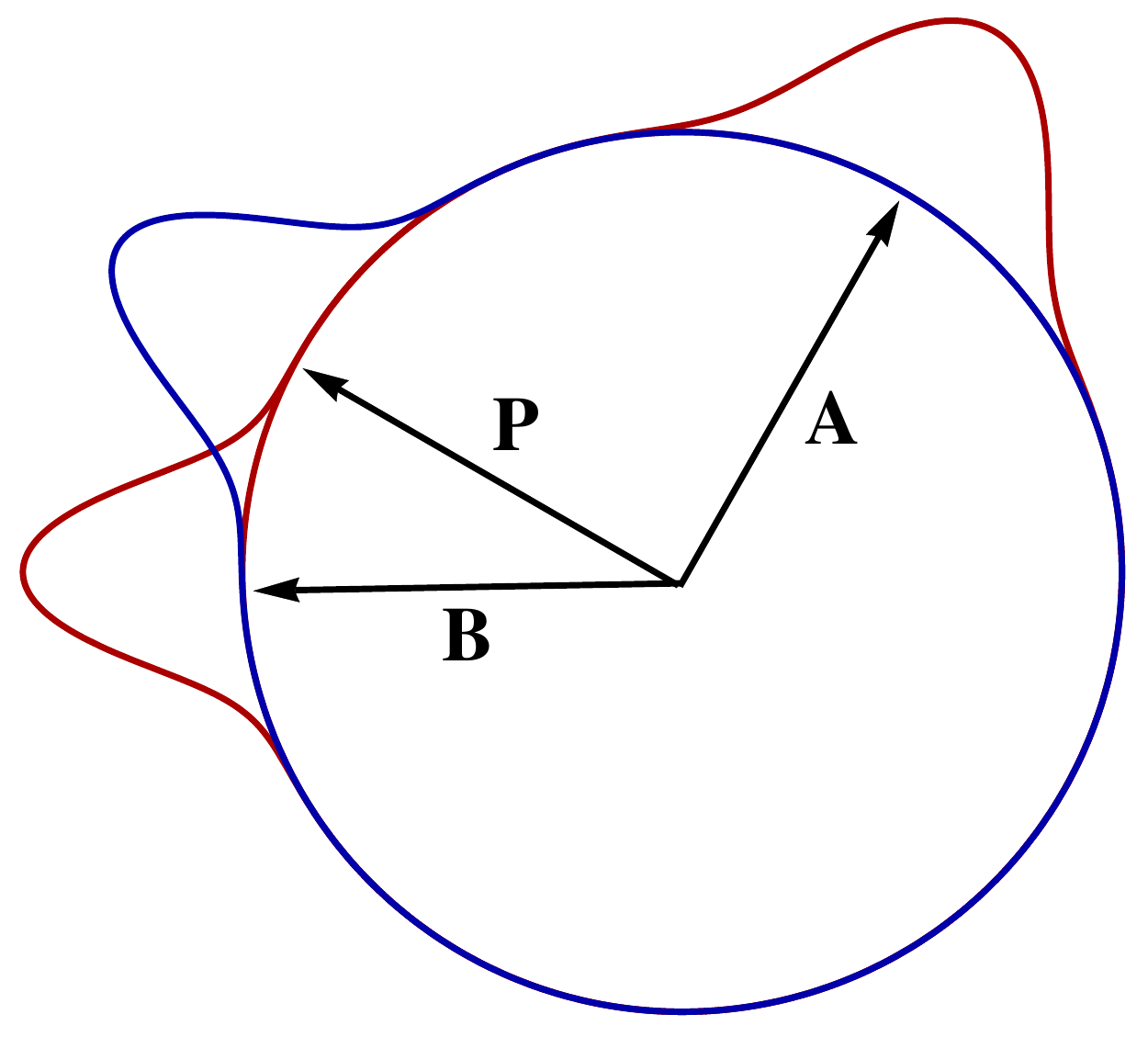}
	\caption{\label{fig:product} An example of the SGF product rule on the unit ring, where $\al = 25$, $\bA = (\cos[\pi/3],\sin[\pi/3])$, $\be = 50$ and $\bB = (\cos\pi,\sin\pi)$ yields $\ze = 25\sqrt{3}$ and $\bP = (\cos[5\pi/6],\sin[5\pi/6])$.}
\end{figure}

%=============================
The normalized ``spherical gaussian function'' (SGF) is
\begin{equation} \label{eq:sgf}
	\Ga(\br) = \frac{\exp(\al \bA \cdot \br)}{\sqrt{2\pi (\pi/\al)^\la I_\la(2\al)}},		\qquad		\br \in \SD
\end{equation}
where $\bA \in \SD$ is a fixed center, $\al \geq 0$ is a fixed exponent, $I_\la$ is a modified Bessel function\cite{NISTbook} and
\begin{equation}
	\la = (D-1)/2
\end{equation}
If we define $\bu = \br - \bA$ then, \alert{for a unit sphere,} we have $u^2 = 2(1 - \bA\cdot\br)$ and $\Ga(\br) \propto \exp\left[\alpha (1 - u^2/2)\right]$ therefore decays as a cartesian Gaussian in $u$.  (See Fig.~\ref{fig:product}.)  
The SGF is single-valued and smooth and decays from a maximum at $\br = \bA$ to a minimum at $\br = -\bA$.  If $\al$ is small, the SGF is almost constant over the sphere;  if $\al$ is large, the SGF is strongly peaked around $\bA$.  For this reason, it is a natural basis function for a localized MO on a sphere.

The product of two SGFs is a third SGF, because
\begin{gather}
	\exp(\al \bA \cdot \br) \exp(\be \bB \cdot \br) = \exp(\ze \bP \cdot \br)	 \label{eq:product}	\\
	\ze = \sqrt{\al^2 + \be^2 + 2\al\be \cos\theta}												\\
	\bP = (\al \bA + \be \bB) / \ze
\end{gather}
where $\cos\theta = \bA \cdot \bB$.  (See Fig.~\ref{fig:product}.)

%================================
\section{Integrals over Spherical Gaussians}
\label{sec:ints}
%================================
The hyperspherical harmonic addition theorem \cite{Avery85} for points on the unit $D$-sphere that subtend an angle $\omega$ is
\begin{equation}
	\Geg = \frac{2\pi}{n+\la} \frac{\pi^\la}{\G(\la)} \sum_K Y_{n,K}^*(\br_1) Y_{n,K}(\br_2)
\end{equation}
where $C_n^\la$ is a Gegenbauer polynomial and $\G$ is the Gamma function.\cite{NISTbook}  The resolution of the Coulomb operator on the $D$-sphere is therefore
\begin{align} \label{eq:RO}
	r_{12}^{-1}	& = (2-2\co)^{-1/2}																						\notag	\\
				& = \sum_{n=0}^\infty \frac{\Braket{(2-2\co)^{-1/2} | \Geg}}{\Braket{\Geg | \Geg}} \Geg					\notag	\\
				& = \frac{4^\la \G(\la)^2}{2\pi} \sum_{n=0}^\infty \frac{\G(n+\half)(n+\la)}{\G(n+\half+2\la)}\ \Geg			\notag	\\
				& = (4\pi)^\la \sum_{n=0}^\infty \frac{\G(n+\half)\ \G(\la)}{\G(n+\half+2\la)} \sum_K Y_{n,K}^*(\br_1) Y_{n,K}(\br_2)
\end{align}

The product rule \eqref{eq:product} yields the overlap integral
\begin{equation} \label{eq:overlap}
	\GGab = \frac{I_\la(\ze)/\ze^\la}{\sqrt{I_\la(2\al) I_\la(2\be) / (4\al\be)^\la}}
\end{equation}
and re-normalized kinetic integral (with $\hat{T} \equiv -\nabla^2/2$)
\begin{equation} \label{eq:kinetic}
%\begin{multline} \label{eq:kinetic}
	\frac{\GTG}{\GGab} = \frac{I_{\la+1}(\ze)}{I_\la(\ze)} \frac{(2\la+1)\al\be\cos\theta}{2\ze}	
	%\\
						- \frac{I_{\la+2}(\ze)}{I_\la(\ze)}\frac{(\al\be\sin\theta)^2}{2\ze^2}
%\end{multline}
\end{equation}
Using the Coulomb resolution \eqref{eq:RO}, it can be shown that the re-normalized electron repulsion integral (ERI), in chemist's notation,\cite{SzaboBook} is
\begin{equation} \label{eq:ERI}
%\begin{multline} \label{eq:ERI}
	\frac{\GGGG}{\GGab \GGcd} = \frac{4^\la \G(\la)^2}{2\pi}	
	%\\
		\times \sum_{n=0}^\infty \frac{\G(n+\half)(n+\la)}{\G(n+\half+2\la)} \frac{I_{n+\la}(\ze)}{I_\la(\ze)} \frac{I_{n+\la}(\eta)}{I_\la(\eta)} C_n^\la(\cos\chi)
\end{equation}
%\end{multline}
where $\eta$ and $\bQ$ are ket analogs of $\ze$ and $\bP$, respectively, and $\cos\chi = \bP \cdot \bQ$.  Special cases of these formulae for $D = 1$ (a ring), $D = 2$ (a normal sphere) and $D = 3$ (a glome) are given in Table \ref{tab:formulae}.  (It should be noted that the ERI for $D=1$ is the finite part of an infinite quantity.\cite{gLDA14})

%========================
\section{Computational Efficiency}
\label{sec:algo}
%========================
In a calculation using $N$ SGFs, computing the non-negligible ERIs is often the most time-consuming step and, for efficiency, one should use both two-center and four-center cutoffs.\cite{Review94}  The Cauchy-Schwarz bound\cite{Haser89, Bound94}
\begin{equation}
	\GGGG \le Z_{\al\be} Z_{\ga\de}
\end{equation}
is particularly useful because the required factors
\begin{align}
	Z_{\al\be}	& = (\Ga\Gb|\Ga\Gb)^{1/2}	\notag	\\
				& = \frac{\GGab}{I_\la(\ze) / \ze^\la} \sqrt{\frac{{_1F_2}(\la+\half,\la+1,2\la+\half,\ze^2)}{2\la\,\G(2\la+\half)\sqrt{\pi}}}
\end{align}
(where $_1F_2$ is the generalized hypergeometric function\cite{NISTbook}) can be found in closed form.  For example, for $D = 2$,
\begin{equation}
	Z_{\al\be} = \frac{\GGab}{i_0(\ze)} \sqrt{\frac{\pi}{2} \frac{\mathbf{L}_0(2\ze)}{2\ze}}
\end{equation}
where $\mathbf{L}_0$ is a modified Struve function.\cite{NISTbook}

In practice, the sum in \eqref{eq:ERI} must be truncated after $M$ terms but this is not problematic because the series converges rapidly.

In summary, we recommend the following algorithm:
\begin{algorithmic}[1]
	\State npairs $\gets 0$ 
	\For {$i = 1,N$}
		\For {$j = i,N$}
			\If {$( G_i | G_j ) >$ threshold}
				\State npairs $\gets$ npairs + 1
				\State Compute $I_{n+\la}(\ze) / I_\la(\ze)$ for $0 \le n \le M$
				\State Compute $T_{ij} = (G_i | \hat{T} | G_j )$
				\State Compute $Z_{ij} = \sqrt{(G_i G_j | G_i G_j)}$
			\EndIf
		\EndFor
	\EndFor
	\For {$ij = 1$,npairs}
		\For {$kl = ij,$npairs}
			\If {$Z_{ij} Z_{kl} >$ threshold}
				\State Compute $(G_i G_j | G_k G_l)$
			\EndIf
		\EndFor
	\EndFor
\end{algorithmic}
The Gegenbauer polynomials needed in step 15 (see Table \ref{tab:formulae}) should be found by forward recursion, \textit{e.g.}
\begin{gather}
	T_n(z) = 2z T_{n-1}(z) - T_{n-2}(z)								\\
	P_n(z) = \frac{2n-1}{n} z P_{n-1}(z) - \frac{n-1}{n} P_{n-2}(z)	\\
	U_n(z) = 2z U_{n-1}(z) - U_{n-2}(z)
\end{gather}

\begin{table*}
	\caption{\label{tab:results}
		Thomson lattices, point groups, vibrational representations $\G_{\rm vib}$, Wigner energies\footnote{$E_0$ is the Coulomb energy of the Thomson lattice; $E_1$ is the harmonic zero-point vibrational energy
		of the lattice.} $E_0$ and $E_1$, optimal single-zeta exponents $\al$, double-zeta HF energies $\EHF$, exact energies $E$ and reduced correlation energies
		$\Ecred$ (all in \mEh) for $n$ same-spin electrons on a 2-sphere with Seitz radius $r_s = 2R/\sqrt{n} = 100$.  The final two rows give the number
		$N_{\rm G}$ of spherical gaussians and number $N_{\rm Y}$ of spherical harmonics required to achieve $\EHF$.}
	\footnotesize
	\begin{ruledtabular}
		\begin{tabular}{cccccccc}	
			$n$					&		2		&		3		&		4		&		6		&		8		&		12		&		24		\\
			Lattice				&	diameter	&	triangle		&  tetrahedron	&  octahedron	&	anti-cube	&  icosahedron	&	snub cube	\\
			Point group			& $D_{\infty h}$	&	$D_{3h}$	&	$T_d$		&	$O_h$		&	$D_{4d}$	&	$I_h$		&	$O$		\\
			$\G_{\rm vib}$		&	$\Pi_u$		&	$A_2''+E'$	&	$E + T_2$	&	$T_{2g} +$					&	$A_1 + B_1 + B_2 +$
																				&	$G_g + H_g +$				&	$2A_1 + 2A_2 +$			\\
								&				&				&				&	$T_{1u} + T_{2u}$			&	$2E_1 + 2E_2 + E_3$
																				&	$T_{1u} + G_u + H_u$		&	$4E + 5T_1 + 6T_2$		\\
			\hline
			$E_0$				&	7.071		&	20.000		&	36.742		&	81.529		&	139.125	&	283.856	&	911.811		\\
			$E_0 + E_1$		&	7.912		&	21.525		&	39.125		&	85.573		&	144.727	&	292.832	&	930.387	\\
			$\al$				&	0.050		&	0.071		&	0.084		&	0.107		&	0.127		&	0.156		&	0.227		\\
			$\EHF$				&	8.263		&	22.194		&	39.822		&	86.438		&	145.929	&	294.256	&	933.275	\\
			$E$				&	7.993		&	21.589		&	39.102		&	---			&	---			&	---			&	---			\\
			$-\Ecred$			&	0.135		&	0.202		&	0.180		&  $\sim 0.14$	&  $\sim 0.15$	&  $\sim 0.12$	&  $\sim 0.12$	\\
			$\nG$				&		2		&		6		&		8		&		12		&		16		&		24		&		48		\\
			$\nY$				&		36		&		36		&		81		&		196		&		144		&	$\ge 225$	&	$\ge 225$	\\
		\end{tabular}
	\end{ruledtabular}
\end{table*}

%====================
\section{Numerical Results}
\label{sec:results}
%====================
In 1904, J.J. Thomson asked\cite{Thomson04} what arrangement of $n$ identical charges on a sphere minimizes their electrostatic energy $E_0$.  This deceptively simple question and its various generalizations have led to much work\cite{Erber97} and, although rigorous mathematical proofs are rare,\cite{Schwartz13} careful numerical investigations\cite{Wales06} have provided optimal or near-optimal arrangements for many values of $n$.

Thirty years later, Wigner discovered\cite{Wigner34} that a low-density electron gas will spontaneously ``crystallize'', each electron moving with small amplitude around a lattice site in what is now called a ``Wigner crystal'' \alert{(or, in case of a finite number of particles, a Wigner molecule)}.  Such crystals have also been observed for electrons confined within harmonic wells, \cite{Cioslowski10, Cioslowski11, Cioslowski12, Cioslowski13a, Cioslowski13b} cubes,\cite{Alavi00} squares\cite{Staroverov10} and spheres.\cite{TEOAS09}

The exact energy of a Wigner molecule can be approximated by the sum of its Thomson energy $E_0$ and the harmonic zero-point energy $E_1$ of the electrons as they vibrate around the lattice sites. \cite{Cioslowski09} These vibrations can be classified according to their irreducible representations $\G_{\rm vib}$ within the point group of the Thomson lattice\cite{SpecBook} (see Table \ref{tab:results}).

To illustrate the usefulness of SGFs, we have studied $n$ same-spin electrons on a 2-sphere with radius $R$ and Wigner-Seitz radius $r_s = R\sqrt{2} = 100$, for seven $n$ values.

We first consider $n = 2$, for which the Thomson lattice is points at the north and south poles of the sphere.  If we place SGFs with exponent $\al$ at each pole and minimize the Hartree-Fock (HF) energy\cite{SzaboBook} with respect to $\al$, we obtain the minimal-basis energy
\begin{equation}
	\EHF^{\al} = 0.008\,270
\end{equation}
Adding a second SGF (with exponent $\be$) at each pole and optimizing with respect to both exponents yields the split-valence energy
\begin{equation}
	\EHF^{\al,\be} = 0.008\,263
\end{equation}
This energy, which is obtained using only $\nG = 4$ SGFs, can also be obtained using a spherical harmonic basis, but only by using harmonics with $0 \le \ell \le 5$, of which there are $\nY = (5+1)^2 = 36$.  This example reveals how much more efficient SGFs are than spherical harmonics, for problems in which the MOs are strongly localized.  It can be shown\cite{TEOAS09} that the exact energy is
\begin{equation}
	E = 0.007\,993
\end{equation}
which implies that the reduced (i.e. per electron) correlation energy\cite{SzaboBook} is $\Ecred = -0.135$ \mEh.

We have performed analogous calculations for all values of $n$ where the Thomson lattice sites are equivalent.  It turns out that there are seven such cases and the results for $n$ = 2, 3, 4, 6, 8, 12, 24 are given in Table \ref{tab:results}.

Although the Wigner-Seitz radius (the average distance between neighboring electrons) is $r_s = 100$ in all cases, we note that the minimal-basis exponent $\al$ grows, i.e.~the electrons become more localized, as $n$ increases.

For $n \ge 6$, we have not been able to calculate the exact energy $E$, so we have estimated the reduced correlation energies in these cases using $E \approx E_0 + E_1$.  The resulting $\Ecred$ values appear to decrease slowly with $n$.

Finally, we note that the superior efficiency of SGFs, compared with spherical harmonics, is observed for all $n$ values that we have considered.  In each case, the number $\nY$ of spherical harmonics required to achieve the HF energy in Table \ref{tab:results} was an order of magnitude larger than the number $\nG$ of SGFs.  In fact, for $n = 12$ and $n = 24$, not even 196 spherical harmonics (i.e.~$0 \le \ell \le 13$) were able to match the energy of the split-valence SGF basis.

%======================
\section{Concluding Remarks}
\label{sec:conclusions}
%======================
Cartesian gaussian basis functions, which are widely used in quantum chemical calculations in $\RD$, can be successfully generalized to spherical gaussian functions (SGFs) for calculations on the sphere $\SD$.  We have derived formulae for the required overlap, kinetic energy and electron repulsion integrals and the worst of these involves a rapidly converging infinite series.

In quantum chemical calculations in $\RD$, it is common to use both $s$-type cartesian gaussians \eqref{eq:gal} and gaussians of higher angular momentum (\textit{i.e.}~$p$-type, $d$-type, etc.).  Integrals over these higher functions can be obtained\cite{Boys50, Review94} from the fundamental integrals over $s$-type functions by differentiating with respect to the cartesian coordinates of the gaussian center.  In a similar way, if desired, one can obtain higher SGFs, and their integrals, by differentiating \eqref{eq:sgf}, \eqref{eq:overlap}, \eqref{eq:kinetic} and \eqref{eq:ERI} with respect to the cartesian coordinates of $\bA$, $\bB$, $\bC$ and/or $\bD$.

We are using SGFs in a systematic study of electrons on 2-spheres and 3-spheres and will report our results elsewhere.

%%%%%%%%%%%%%
\begin{acknowledgments}
%%%%%%%%%%%%%
P.F.L.~and P.M.W.G.~thank the NCI National Facility for generous grants of supercomputer time.
P.M.W.G.~thanks the Australian Research Council for funding (Grants No.~DP120104740 and DP140104071). 
P.F.L. thanks the Australian Research Council for a Discovery Early Career Researcher Award (Grant No.~DE130101441) and a Discovery Project grant (DP140104071).
%%%%%%%%%%%%%
\end{acknowledgments}
%%%%%%%%%%%%%

\bibliography{SphGauss}

\end{document}